\newif\ifreport
\reporttrue

\newif\ifdraft
\drafttrue

\ifreport
\documentclass[12pt]{article}
\usepackage{graphicx,latexsym}
\else

\ifdraft
\documentclass[times, 12pt, pre,aps, preprint]{revtex}
\draft
\else
\documentclass[times, 12pt,pre,aps,twocolumn]{revtex}
\fi
\usepackage{graphicx}

\fi

\relax 
\def\figureI#1{
\begin{figure}[#1]
\centering
\includegraphics[scale=\figscale]{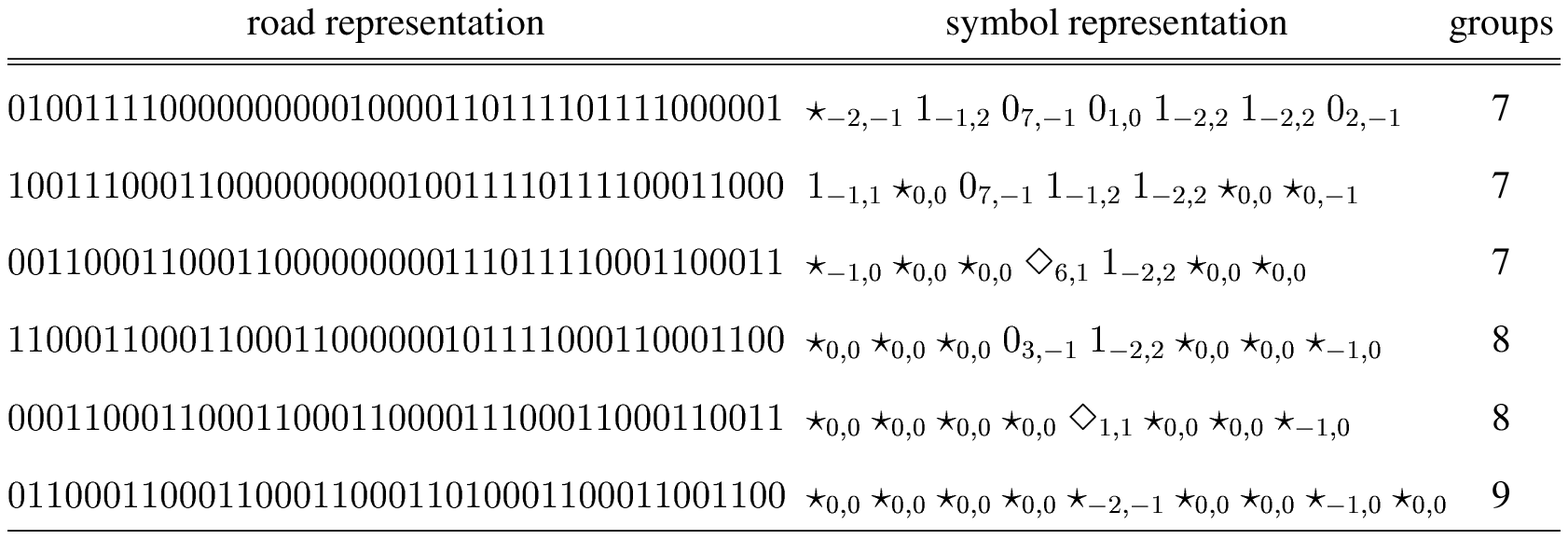}
\caption{
\label{fig-evolution}
An example time evolution under rule $\Rule 3,2$
from an initial state with $L=41$ into a cyclic state,
expressed both in the road representation and the symbolic representation
introduced in Sec.~\ref{sec-symbol-def} in the text.
Periodic boundary conditions are used.
}
\end{figure}
}

\def\figureII#1{
\begin{figure}[#1]
\centering
\includegraphics[scale=\figscale]{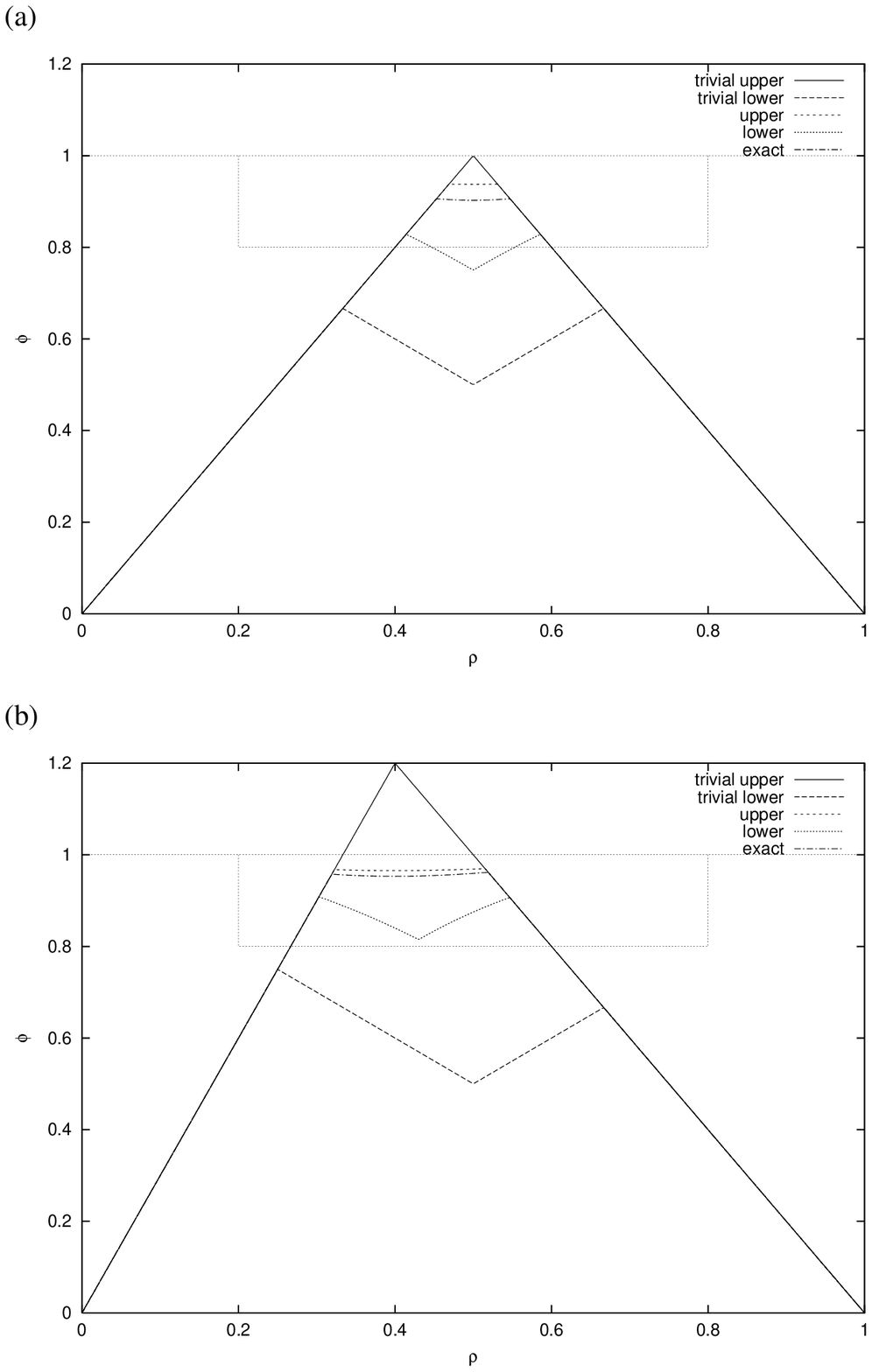}
\caption{\label{fig-limits}
Different limits and the exact solution
for the flow $\phi$ as a function of density $\rho$
at infinite length for (a) $\Rule 2,2$ and (b) $\Rule 3,2$.
The trivial limits are given by Eqs.~(\ref{eq-trivial-min}) and
(\ref{eq-trivial-max}).
The equations (\ref{eq-infinite-max}) and
(\ref{eq-infinite-min}) yield the tighter limits.
The enclosed regions are shown magnified in Fig.~\ref{fig-infinite}.
}
\end{figure}
}

\def\figureIII#1{
\begin{figure}[#1]
\centering
\includegraphics[scale=\figscale]{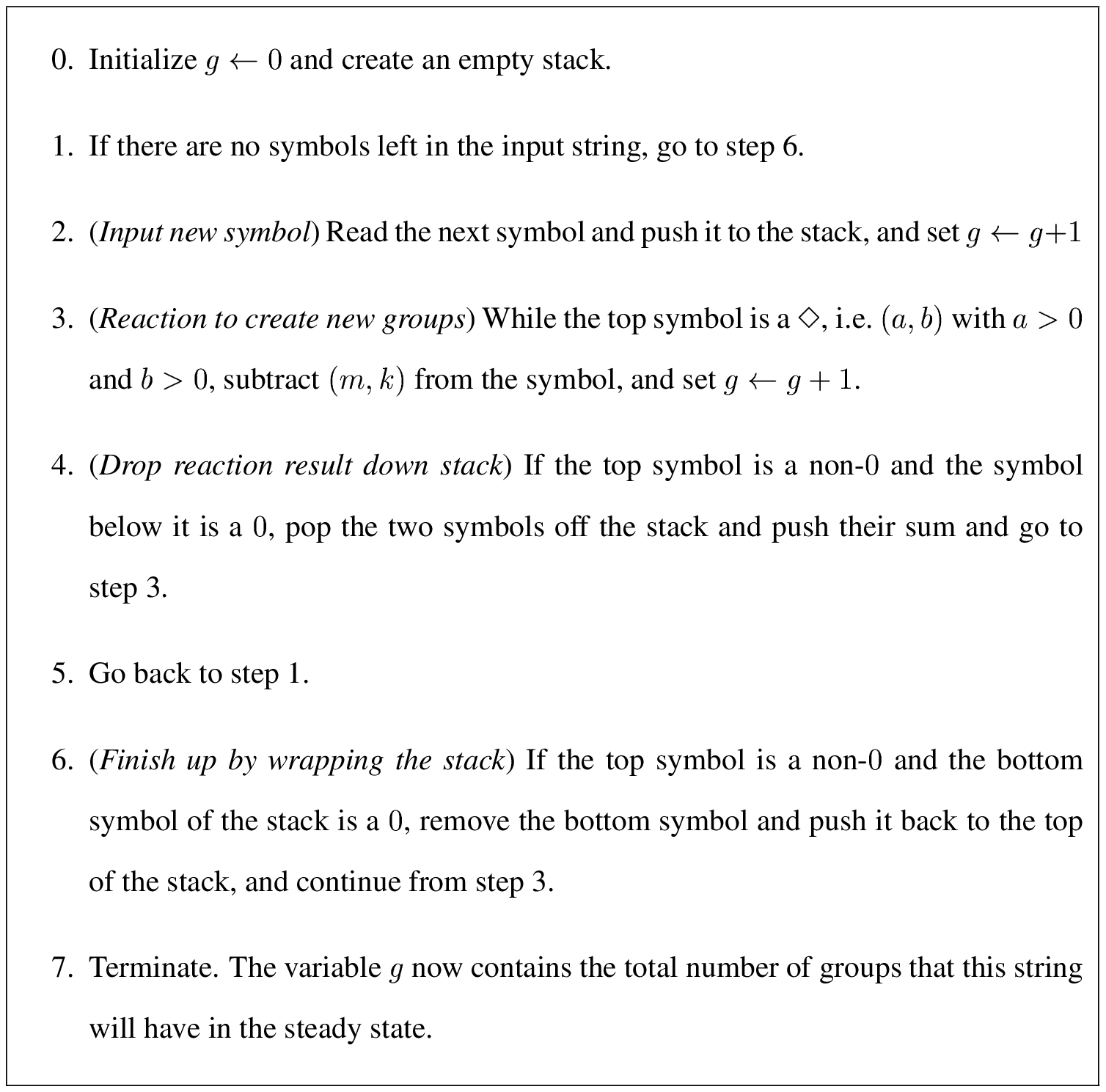}
\caption{
\label{fig-stackalg}
This algorithm calculates the final number of groups in the
given symbol string with periodic boundary conditions.
The symbols $?_{a,b}$ are represented as pairs $(a,b)$ and
the input string is read left to right.
Addition between pairs is defined as
$(a,b)+(c,d)=(a+c, b+d)$.
Step~\ref{item-alg-react} corresponds to
Eq.~(\ref{eq-diamondreact}) and
steps~\ref{item-alg-drop} and \ref{item-alg-finish} both correspond to
Eqs.~(\ref{eq-zerostarreact}) and (\ref{eq-zeroonereact})
in the text.
}
\end{figure}
}

\def\figureIV#1{
\begin{figure}[#1]
\centering
\includegraphics[scale=\figscale]{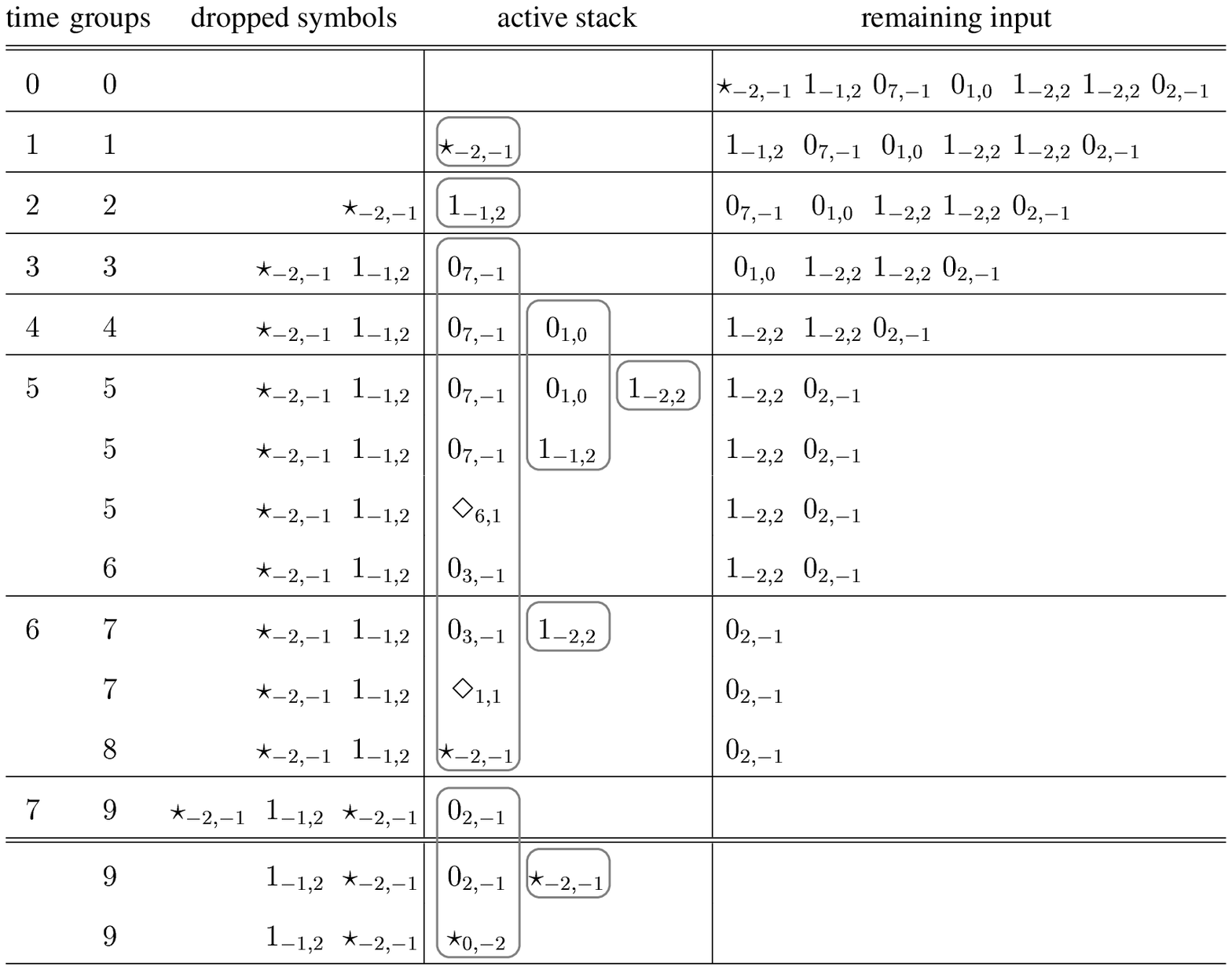}
\caption{
\label{fig-evolution-stack}
Evolution of the stack of the algorithm in Fig.~\ref{fig-stackalg} operating
on the sample string in Fig.~\ref{fig-evolution}.
In the end, the number of extra groups is the same as in
Fig.~\ref{fig-evolution}.
The doubled line signifies the end of the input string, after which
the stack is wrapped over (cf. step~\ref{item-alg-finish} of the
algorithm in Fig.~\ref{fig-stackalg}).
The division to dropped symbols and active stack is implicit --
it is only significant for the Markov model analysis of the algorithm
in Section~\ref{sec-exact-solution}.
The boxes depict the ``lifetimes'' of the sub-stacks started
by each input symbol.
The time steps of the Markov model are at the bottom lines of
these boxes, where the sub-stacks are finished.
}
\end{figure}
}

\def\figureV#1{
\begin{figure}[#1]
\centering
\includegraphics[scale=\figscale]{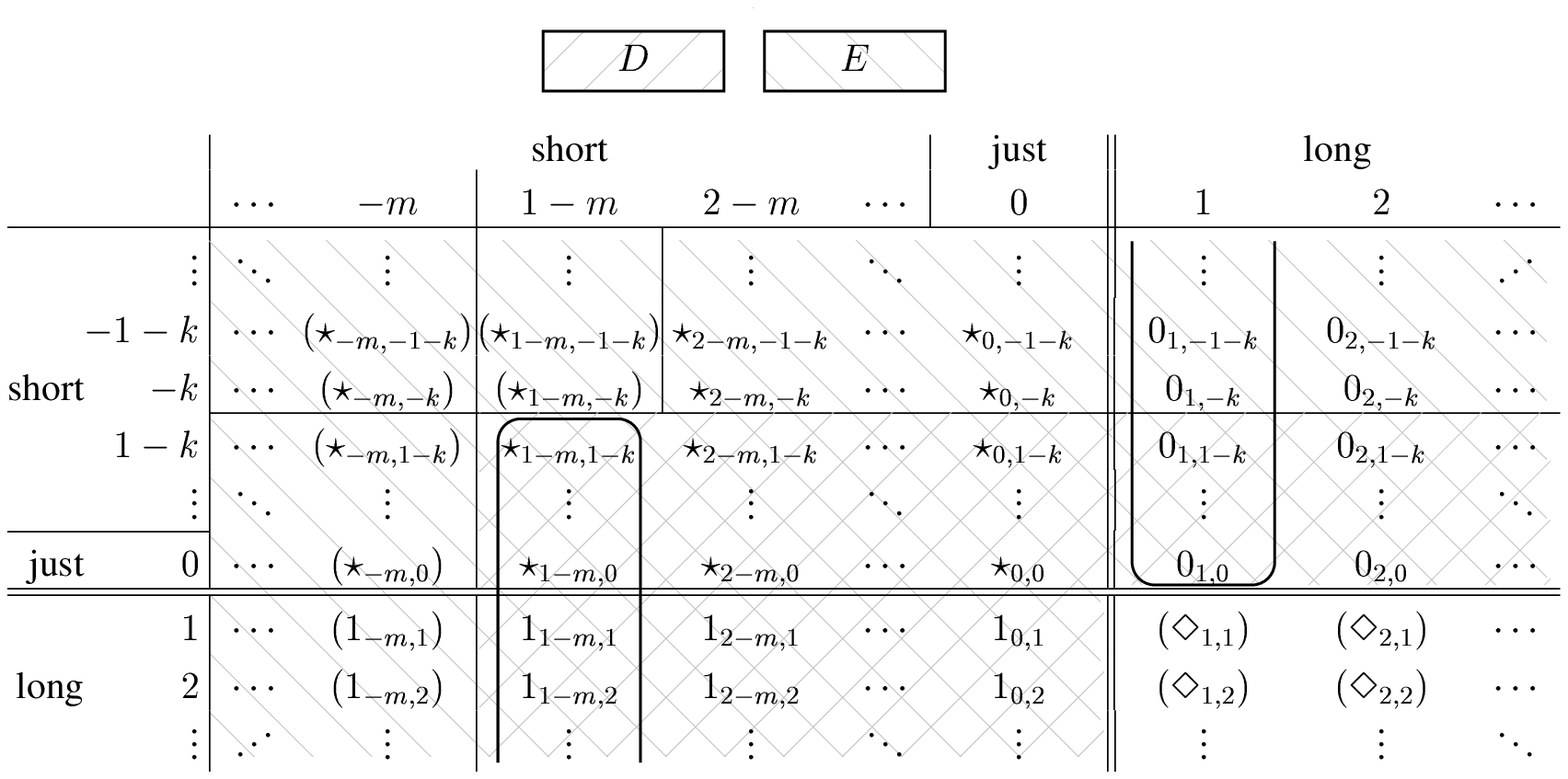}
\caption{\label{fig-symbols} The possible index combinations for
different symbols in the Markov model for the algorithm
discussed in Sec.~\ref{sec-exact-solution}.
By considering the possible outcomes of combined $0$ and non-$0$ symbols,
it can be seen that the parenthesized symbols can not occur.
The sets $D$ and $E$ defined in Sec.~\ref{sec-exact-solution}
are depicted by the different shadings. The marked sequences
are represented by the generating functions $\G_{1-m}(z)$ and $\G_1(z)$
discussed in Sec.~\ref{sec-genfunc}.
}
\end{figure}
}

\def\figureVI#1{
\begin{figure}[#1]
\centering
\includegraphics[scale=\figscale]{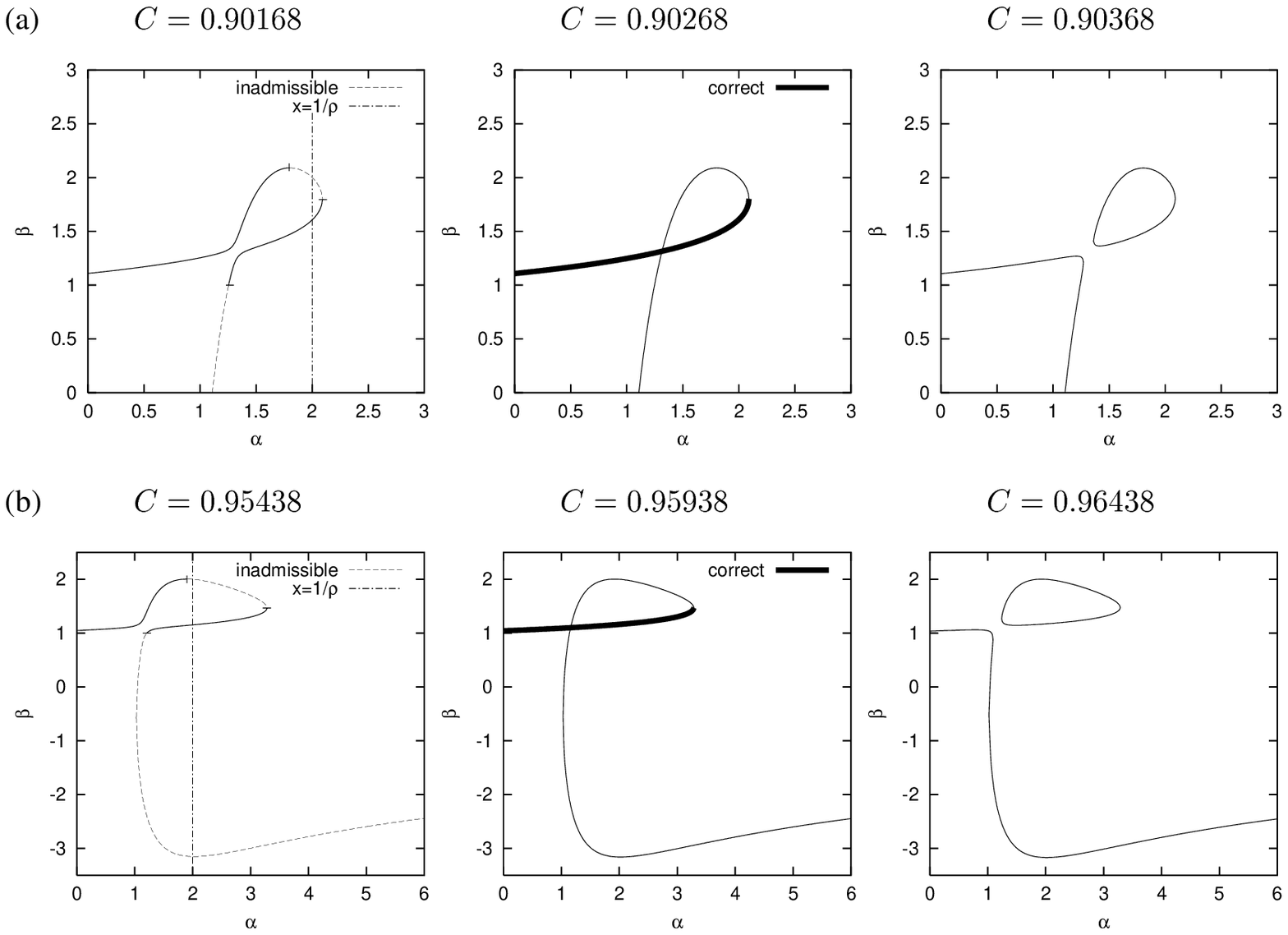}
\caption{\label{fig-g-loop}
Plots of the real solutions $\beta = \G_1(1/\alpha) + 1$ of the generating
function equation (\ref{eq-g-symmetric}) at $\rho=1/2$ for $C$
below, at, and above the correct $\phi$ for
(a) $\Rule 2,2$ and (b) $\Rule 3,2$.
Because $\beta$ is an ordinary generating function for a sequence of
positive values, the correct solution and all its derivatives must
be positive.
Furthermore, because $\alpha = 1/\rho$ corresponds to a sum of
probabilities, $\beta$ must converge at least for $|\alpha|\le1/\rho$.
This excludes too large values of $C$. For too small values,
the lower curve continues below $1$ yielding negative probabilities
and the upper curve is not feasible either,
because it is decreasing at $\alpha = 1/\rho$. Only the singular curve at the correct $C$
yields an admissible solution.
Because Eq.~(\ref{eq-g-symmetric}) is symmetric with respect to $(\alpha,k)$ and $(\beta,m)$,
the solutions have symmetry axis $\alpha=\beta$ when $m=k$. }
\end{figure}
}

\def\figureVII#1{
\begin{figure}[#1]
\centering
\includegraphics[scale=\figscale]{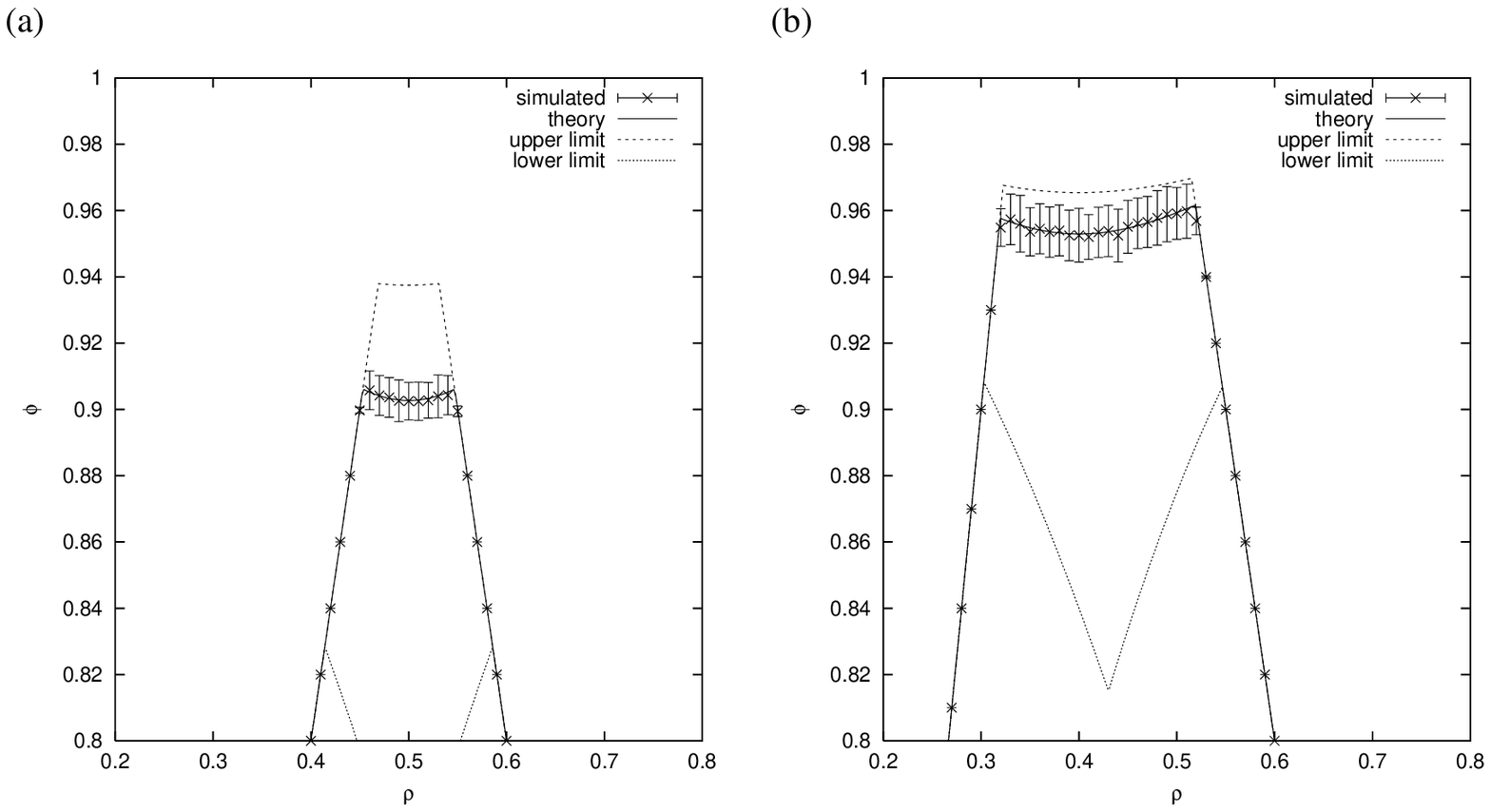}
\caption{\label{fig-infinite}
Infinite-space behavior of flow as a function of density under (a) $\Rule 2,2$ and (b) $\Rule 3,2$.
Average of 100 simulations with $L=10000$, theoretical flow,
and upper and lower limits given by
Eqs.~(\ref{eq-infinite-max}) and (\ref{eq-infinite-min}).
}
\end{figure}
}

\def\figureVIII#1{
\begin{figure}[#1]
\centering
\includegraphics[scale=\figscale]{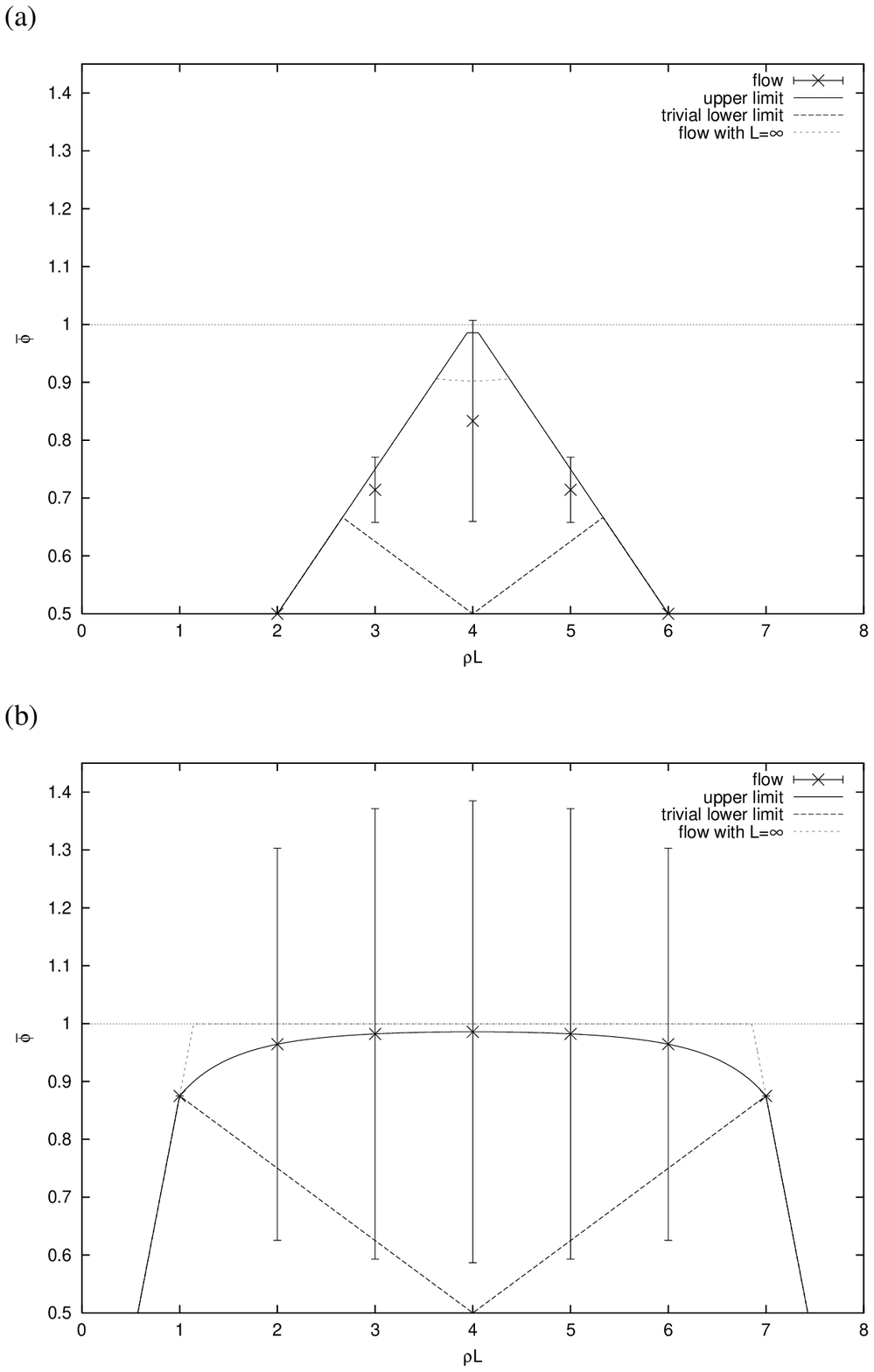}
\caption{\label{fig-finite}
Finite-space behavior of the average flow $\overline \phi$ as a function of density
under (a) $\Rule 2,2$ and (b) $\Rule 7,7$ when $L=8$.
All points represent either evaluated formulas or averages calculated
over all configurations.
The trivial lower limit is given by Eq.~(\ref{eq-trivial-min}).
The upper limit given by Prop.~\ref{prop-finite-max} is exact for $m,k\ge L-1$
as demonstrated by b).
}
\end{figure}
}

\begin{document}

\def\adr{
University of Jyv\"askyl\"a\\
Department of Mathematical Information Technology\\
P.O.~Box~35 (Agora), FIN-40351~Jyv\"askyl\"a, FINLAND}

\ifreport

\ifdraft\else
\input{report7_2001_frontmatter.tex}
\fi

\title{Exact limiting solutions for certain\\deterministic traffic rules}
\author{Janne V. Kujala and Tuomas J. Lukka~%
\footnote{E-mail addresses: jvk@iki.fi, lukka@iki.fi}\\\\\adr}
\date{} %

\def\text#1{\mbox #1}
\fi

\maketitle

\def\figscale{1}
\ifdraft\def\figscale{0.94}\fi
\ifreport\def\figscale{0.9}\fi

\def\Rule#1,#2{{\mathcal R}_{#1,#2}}
\def\Rulef#1,#2{{\mathcal F}_{#1,#2}}
\def\Rules#1,#2{{\mathcal S}_{#1,#2}}
\def\Ruleflet{{\mathcal F}}
\def\Ruleslet{{\mathcal S}}
\def\g{f}
\def\G{F}

\newtheorem{prop}{Proposition}
\newtheorem{coro}{Corollary}[prop]

\begin{abstract}
We analyze the steady-state flow as a function of the initial density for
a class of deterministic cellular automata rules (``traffic rules'')
with periodic boundary conditions [H. Fuk\'s and N. Boccara,
Int.~J.~Mod.~Phys.~C {\bf 9}, 1 (1998)].
We are able to predict from simple considerations
the observed, unexpected cutoff of
the average flow at unity.
We also present an efficient algorithm for determining the exact final
flow from a given finite initial state. We analyze the behavior of
this algorithm in the infinite limit to obtain for $\Rule m,k$
an exact polynomial equation maximally of $2(m+k)$th degree in the flow
and density.
\ifreport

\medskip\noindent
{\bf Keywords:} Cellular automata, traffic modeling, generating functions\\
{\bf AMS Subject Classification:} 37B15, 68Q80\\
{\bf PACS:} 45.70.Vn
\fi
\end{abstract}

\section{Introduction}

There is considerable interest in modeling traffic behavior
via one-dimensional cellular automata (CAs). The original models by
Nagel and Schreckenberg\cite{nagel-schreckenberg} and
Fukui and Ishibashi\cite{fukui-ishibashi} are analyzed in
e.g.~\cite{wang-statistical,%
wang-analytical,%
fuks-exact},
and the more general behavior
of sum-conserving CAs is considered in \cite{%
boccarafuks-numberconserving,%
boccarafuks-conserving%
}.
In \cite{fuksboccara-generalized},
Fuk\'s and Boccara introduced an interesting
class of generalized deterministic traffic rules $\Rule m,k$, which display
a surprising steady-state
behavior:
the expected flow of the cars never exceeds one
regardless of the constraint values $m$ and $k$.

In these rules, as is usual for traffic rules, the road is represented
as a one-dimensional lattice where each site has as its value either
$0$ (empty) or $1$ (car).
Under $\Rule m,k$,
a block of cars (ones)
at most $k$ units long moves right at most $m$ units, or to the beginning
of the next group.
The same rule can also be expressed
as follows: at each turn, each maximal match of $1^x0^y$ is
replaced (see Fig.~\ref{fig-evolution}):
$$ 1^x0^y \rightarrow 1^{x-a}0^b1^a0^{y-b},$$
where $a=\min\{k,x\}$ and $b=\min\{m,y\}$.
From this representation, the dualism
between the motion of the cars under the rule $\Rule m,k$ and
the motion of the empty sites under rule $\Rule k,m$ in the opposite
direction, as mentioned
in \cite{fuksboccara-generalized}, is obvious.

\ifreport
\figureI{tbp}
\fi

The ``physical'' quantities of interest
in systems that obey these rules are $\rho$, the
density of ones, and $\phi$, the flow, defined as $\rho \left<v\right>$,
where $\left<v\right>$ is the average velocity of the cars.
For finite-length systems,
we write $\phi$ for the time-averaged steady-state flow from a single
state and $\overline \phi$ for the average of $\phi$ over all states.
For infinite-length systems, $\phi$ is the steady-state flow.
The equation
\begin{equation}
\label{eq-dualism}
\overline \phi_{\Rule m,k}(\rho) =
\overline \phi_{\Rule k,m}(1-\rho)
\end{equation}
expresses one consequence of the dualism discussed above.
There are also other quantities such as acceleration,
but these are outside the scope of this article.

In this article, we examine the steady-state
flow of $\Rule m,k$, obtaining an exact polynomial equation
in the infinite case.
In the following sections, we first develop a formalism based on
representing the road as a sequence of blocks
rather than single sites.
We show that the average flow is fully determined by the number
of these blocks in the steady state.
In Sections~\ref{sec-simple-limits} and \ref{sec-algorithm},
we use this fact to obtain
simple upper and lower limits for the average flow and
present an efficient algorithm for calculating
the steady-state flow from a given finite initial state.
In Section~\ref{sec-exact-solution},
we consider the behavior of the algorithm in the infinite limit
and derive a steady-state condition,
which we then solve in Section~\ref{sec-genfunc},
yielding
an analytical solution
in the case of an infinite space.
Finally, in Section~\ref{sec-finite}, we obtain
a non-trivial upper limit for the expected average flow in a finite space.

\section{Fundamental properties of
\mbox{$\Rule {\lowercase{m}},{\lowercase{k}}$}}
\label{sec-fundamental}

The flow of cars
under rule $\Rule m,k$ is easier to understand
if the state
of the road
is considered as a sequence of blocks
instead of single cars.
As we shall see later, it is practical to distinguish between
{\em short}, {\em just}, and {\em long} blocks,
comparing the length of
a block with $m$ or $k$ as follows:
a block of zeroes less than $m$ sites long is a short block,
more than $m$ sites long is a long block,
and exactly $m$ sites long is a just block.
For blocks of ones, the length is compared with $k$ in a similar fashion.
We say that a pair of a $0$-block and a $1$-block is a {\em group}
and define $\rho_G$ as the density of these groups.
We also define
$\rho_{0+}$ and $\rho_{0-}$ as the densities of long and
short $0$-blocks, respectively, and similar symbols for the $1$-blocks.

The states of the system can be divided into nine different categories
by the existence of short and long blocks, see Table~\ref{tab-states}.
In the following, we will consider the three cyclic types of states
separately.
To verify that these are the only cyclic states we first show
that the following two cases are unstable:
long and short blocks of one kind, $\rho_{1+}>0$ and $\rho_{1-}>0$,
and
long blocks of both kinds, $\rho_{1+}>0$ and $\rho_{0+}>0$.
The other cases follow from the same proofs by duality.

\ifreport
\begin{table}
\centering
\footnotesize
\begin{tabular}{lccc}
\hline\hline
Lengths & $\rho_{0-}\ge0$, $\rho_{0+}=0$ & $\rho_{0-}>0$, $\rho_{0+} > 0 $ &
$\rho_{0-}=0$, $\rho_{0+}>0$ \\
\hline
$\rho_{1-}\ge0$, $\rho_{1+}=0$ &
Cyclic: Intermediate & Uncyclic & Cyclic: Free-flowing \\
$\rho_{1-}>0$, $\rho_{1+} > 0 $ &
Uncyclic & Uncyclic $\rho_G$ may increase &
Uncyclic, $\rho_G$ may increase \\
$\rho_{1-}=0$, $\rho_{1+}>0$ & Cyclic: Congested &
Uncyclic, $\rho_G$ may increase &
Uncyclic, $\rho_G$ may increase \\
\hline\hline
\end{tabular}
\caption{
\label{tab-states}
The different types of states in the system. As discussed in the text,
the existence of short and long blocks distinguishes the different types
of states.
}
\end{table}
\fi

First, we note that new long blocks can never form, because a non-long block
of cars moves continually to the right and therefore can not absorb
other cars from the left. This also implies that a non-long block
can only be absorbed to an already long block on the right.
The duality proves the same for long $0$-blocks.
Furthermore, the length of a long block can not grow, because
long blocks emit just blocks whereas they can only absorb short
or just blocks.

To show that the states that have both short and long
blocks of one type are unstable,
consider the sequence
$$
1^x(0^{\le m}1^k)^z0^{\le m}1^y,
$$
where $x<k$, $y>k$, and $0^{\le m}$ represents a $0$-block that is either short
or just. On each of the $z$ first steps,
the block $1^y$ absorbs one just block and emits
one just block from the other end -- its length remains unchanged. However,
on the $(z+1)$th step it absorbs a block of length $x$
and emits a block of length
$k$. Therefore, the number of short blocks has decreased by one, and the length
of the long block has decreased by $k-x$, possibly transforming it into
a just or short block. Mathematical induction using this argument
shows that, if all $0$-blocks are short or just, then during the simulation,
either the number of short or long $1$-blocks drops to zero.
The number of groups does not change in this process.
An important observation is that
the long block behaves like a decaying quasi-particle
that is moving in the opposite direction from the ones
by continuously absorbing short or just blocks
and emitting just blocks.
Naturally, applying the dualism property proves the same for long and short
$0$-blocks.

Next, we show that if there are both long $1$-blocks and long $0$-blocks,
the state is unstable.
We can first apply the
above property to show that either a long $1$-block decays
or it eventually meets
a long $0$-block moving in the opposite direction
(or else there are no long $0$-blocks
left in the system).
But when the long blocks meet, they react and annihilate each other partially
or wholly: the group
\begin{equation}\label{eq-diamondblock}
0^x 1^y,
\end{equation}
where $x>m$ and $y>k$ emits just $0$-blocks leftwards
and just $1$-blocks rightwards
at each time step, reducing to
\begin{equation}\label{eq-diamondresult}
0^{x-m} 1^{y-k}
\end{equation}
and increasing the number of groups by one.
Therefore, eventually either the long $0$-blocks,
the long $1$-blocks or both will
be exhausted, which shows that the initial state is not cyclic.

This completes the study of the uncyclic states,
showing that from any initial state we will finally end up in one
of the remaining states shown in Table~\ref{tab-stable}.
We discuss these three cyclic types of states separately below.

\ifreport
\begin{table}
\centering
\begin{tabular}{lcl}
\hline\hline
Description & Conditions & $\phi$ \\
\hline
Free-flowing & $\rho_{1+} = 0, \qquad \rho_{0-} = 0$ & $ m \rho $ \\
Intermediate & $\rho_{1+} = 0, \qquad \rho_{0+} = 0$ &
$ \rho (1-\rho) / \rho_G$ \\
Congested & $\rho_{1-} = 0, \qquad \rho_{0+} = 0$ & $ k (1-\rho) $ \\
\hline\hline
\end{tabular}
\caption{
\label{tab-stable}
The three different types of cyclic states for $\Rule m,k$ and the formula
for the flow $\phi$
in each.
}
\end{table}
\fi

It is fairly easy to see that
if there are only short and just blocks,
then the $1$-blocks move in the positive direction
and the $0$-blocks move in the negative direction but
the number of blocks and the distribution
and relative order of the $0$ and $1$-blocks
among themselves do not change -- the state is obviously cyclic.
To evaluate the flow, we first note that
in such a state
each block of cars travels
on each step on average $(1-\rho) / \rho_G$ units forwards,
which, when multiplied by the density of cars yields
\begin{equation}
\phi = {\rho (1-\rho) \over \rho_G},\qquad\rho_{0+}=0,\ \rho_{1+}=0.
\end{equation}
As this class of states does not correspond
to either the free-flowing or congested phases of the
simpler traffic rules $\Rule 1,k$ and $\Rule m,1$,
we term it, for want of a better name, intermediate.

The free-flowing states where
all $0$-blocks are long or just and all $1$-blocks are short or just
are also simple.
All the cars obviously move forwards at maximum speed and consequently,
these states are also cyclic, with
\begin{equation}
\phi = m \rho,\qquad\rho_{0-}=0,\ \rho_{1+}=0.
\end{equation}
Applying the dualism
between zeroes and ones
we obtain the formula
\begin{equation}
\phi = k (1-\rho),\qquad\rho_{0+}=0,\ \rho_{1-}=0,
\end{equation}
for the opposite case: congested states with
long or just $1$-blocks and short or just $0$-blocks.

We can summarize the above by noting that
$\rho$ and the final $\rho_G$ determine the type of the cyclic state.
This follows from the fact that the final average lengths
of $0$- and $1$-blocks
can hold for only one type of a cyclic state:
in the intermediate phase blocks must on the average be short or just
whereas in the free-flowing and congested phases either
$0$- or $1$-blocks must be long and the other blocks short or just
as shown in Table~\ref{tab-stable}.
Writing these conditions in terms of $\rho$ and $\rho_G$
allows us to combine
the above evaluations of $\phi$ into one surprisingly
simple formula:

\begin{prop}
\label{prop-oneflow}
The average flow over a cycle
in any cyclic state of $\Rule m,k$ %
is
$$
\phi = \min \bigl\{m\rho,\, \rho (1-\rho) / \rho_G, \, k (1-\rho) \bigr\}.
$$
\end{prop}

The treatment of long blocks above also yields the following
proposition.

\begin{prop}
\label{prop-rhoginc}
During evolution of the system, $\rho_G$ can never decrease. It can
increase only when $\rho_{0+}>0$ and $\rho_{1+}>0$.
\end{prop}

These two propositions give the system its interesting
characteristics: the final flow is completely determined by the
final value of $\rho_G$,
which in turn depends on the intricate reactions
of the long blocks.

As a direct consequence of these propositions,
it is straightforward to obtain crude upper and lower limits
under both finite and infinite length (see Fig.~\ref{fig-limits}):

\begin{prop}
The flow of any cyclic state (and thus the average flow over different
states) satisfies
\begin{eqnarray}
\label{eq-trivial-min}
\phi & \ge &
\min\{ m\rho,
\left| \rho- 1/2 \right| + 1/2 ,
k(1-\rho)
\} \\
\label{eq-trivial-max}
\phi & \le & \min\{m\rho, k(1-\rho)\}
\end{eqnarray}
for any $0\le\rho\le1$ and any size of the lattice.
\end{prop}

The lower limit is obtained through considering what
is the greatest possible number of groups.
For a lattice of size $L$, this is easily seen
to be
$$
N_{G, \max} = L\left( \frac12 - \left|
\rho - \frac12 \right| \right),
$$
which together with Prop.~\ref{prop-oneflow}
yields the first part. The second inequality follows directly
from Prop.~\ref{prop-oneflow}.

When $m=1$ or $k=1$, this proposition reduces to the well-known
flow formula
$$
\phi = \min \bigl\{ m\rho,\, k(1-\rho) \bigr\}
$$
discussed e.g.~in \cite{fuksboccara-generalized}.

\ifreport
\figureII{tbp}
\fi

\section{Simple upper and lower limits at infinite length}
\label{sec-simple-limits}

At infinite length, we can use block probabilities
\begin{equation}
p(a_1a_2\cdots a_n) = \rho^{\sum_i [a_i=1]}(1-\rho)^{\sum_i [a_i=0]}
\end{equation}
in the initial random state
in order to calculate various statistics.
The brackets $[\cdots]$ represent Iverson's notation\cite[p.~11]{iverson},
which evaluates to $1$ if the enclosed statement is true and to $0$
otherwise.

The frequency of groups in the initial, random state
is obviously given by the density of group edges:
$$
\rho_{G,\text{initial}} = p(01) = \rho (1-\rho).
$$
Proposition~\ref{prop-rhoginc} tells us that
$$\rho_G > \rho_{G,\text{initial}}.$$
Combining these with Proposition~\ref{prop-oneflow}
yields the important upper limit
\begin{equation}
\phi \le \min\left\{ m\rho, 1, k(1-\rho)\right\},
\end{equation}
which is valid for {\em any} $\Rule m,k$, as Fuk\'s and Boccara
observed experimentally in
\cite{fuksboccara-generalized}.
The expected flow is cut off by the fact that
initial groups can never combine into fewer groups, which
limits the maximum speed of cars.
We shall show in Section \ref{sec-finite}
that this limit is also valid for finite-length systems.

This limit does not take into account
the dynamics of the system but assumes that the final state has
the same number of groups as the initial state.
It is possible
to improve this upper limit by explicitly including some cases where
certain initial configurations are known to produce new groups.
For adjacent long blocks of zeroes and ones that directly
react with each other as shown by Eqs.~(\ref{eq-diamondblock}) and
(\ref{eq-diamondresult}), we have
\begin{eqnarray*}
\rho_G &\ge& \sum_{x,y\ge1}
\min\left\{\left\lceil{x\over m}\right\rceil,
\left\lceil{y\over k}\right\rceil\right\}
p(1 0^x 1^y 0) \\
& = & \sum_{r\ge0} p(0^{mr+1} 1^{kr+1}) =
{\rho (1-\rho) \over 1 - \rho^k (1-\rho)^m },
\end{eqnarray*}
which yields the tighter upper limit
\begin{equation}\label{eq-infinite-max}
\phi \le \min\left\{ m\rho, 1-\rho^k(1-\rho)^m, k(1-\rho) \right\}
\end{equation}
when combined with Proposition~\ref{prop-oneflow}.

A lower limit can be estimated by computing the density of groups
that could arise by having all long blocks split maximally.
For ones, this is
$$
\rho_G \le \sum_{x\ge1} \left\lceil{x\over k}\right\rceil p(0 1^x 0) =
\sum_{r\ge0} p(0 1^{kr+1}) =
{\rho (1-\rho) \over 1-\rho^k}.
$$
Combining with the same formula for zeroes yields
\begin{equation}\label{eq-infinite-min}
\phi \ge \min\biggl\{m\rho, \max\bigl\{1-\rho^k,
1-(1-\rho)^m\bigr\}, k(1-\rho)
\biggr\}.
\end{equation}

Figure~\ref{fig-limits} shows the accuracy of these limits for
$\Rule 2,2$ and $\Rule 3,2$. For larger $m$ and $k$, the limits
become tighter, converging to unity geometrically as
$m$ and $k$ tend to infinity.

\section{Efficient algorithm for
computing\\\protect\nopagebreak the steady-state flow}
\label{sec-algorithm}

It is not necessary to carry out the simulation to find
the number (or density) of groups in the final state, and thus the
steady-state flow.
In this section, we present an $O(L)$ algorithm for
finding
the final number of groups
from a given starting state
with periodic boundary conditions in a lattice of length $L$.
This is interesting from two different perspectives: first,
it makes it possible to calculate
the final flow for long strings more effectively.
Additionally, it helps
us to understand the dynamics of the system and
derive analytic results on the behavior of the system in
the next Sections.

Let us define the symbols \label{sec-symbol-def}
$\star_{a,b}$, $\Diamond_{a,b}$, $0_{a,b}$ and $1_{a,b}$ all to correspond
to $0^{a+k} 1^{b+m}$ for different $a$ and $b$:
\begin{equation}
0^{a+k} 1^{b+m} = \left\{
\begin{array}{l}
\star_{a,b} \mbox{, $a \le 0$ and $b \le 0$, } \\
0_{a,b} \mbox{, $a > 0$ and $b \le 0$, } \\
1_{a,b} \mbox{, $a \le 0$ and $b > 0$, } \\
\Diamond_{a,b} \mbox{, $a > 0$ and $b > 0$. }
\end{array}
\right.
\label{eq-defsyms}
\end{equation}
We represent the initial state in this notation and then
carry out a series of string replacements.
The final number of groups is obtained as the initial
number of symbols plus the number of extra groups created
by the replacements as shown below.

Diamonds react as follows:
\begin{equation}
\label{eq-firstreact}\label{eq-diamondreact}
\Diamond_{a,b} \rightarrow ?_{a-m, b-k}, \qquad \Delta g = +1,
\end{equation}
where the wildcard symbol $?$
represents the correct symbol for the new indices from
Eq.~(\ref{eq-defsyms}) and
$\Delta g$ represents the number of new groups created by the reaction.

Stars can be collected, along with zeroes and ones,
but zeroes only from the right and ones only from the left.
No new groups are created.
\begin{eqnarray}
\label{eq-firstcombine}%
\star_{a,b} \star_{c,d} & \rightarrow & \star_{a+c, b+d},
\qquad \Delta g = 0, \\
\label{eq-zerostarreact}%
\label{eq-firstdecay}%
0_{a,b} \star_{c,d} & \rightarrow & ?_{a+c, b+d},
\qquad \Delta g = 0, \\
\label{eq-staronereact}%
\star_{a,b} 1_{c,d} & \rightarrow & ?_{a+c, b+d},
\qquad \Delta g = 0.
\end{eqnarray}
Finally, when a $0$ and a $1$ meet, it is possible to form new
groups:
\begin{equation}
\label{eq-lastreact}\label{eq-zeroonereact}
0_{a,b} 1_{c,d} \rightarrow ?_{a+c, b+d},
\qquad \Delta g = 0. \\
\end{equation}
Note that new groups are not created directly by this reaction,
but the result may be a $\Diamond$.%

The above replacements
recursively evaluate the reactions between long blocks.
To prove the correctness of a maximal application of
the replacements,
we consider the reactions occurring in an actual simulation.
As described in Section~\ref{sec-fundamental},
all new groups arise from the annihilations of long blocks
and the length of a long block can never grow.
Thus, the number
of extra groups is dictated by the sequence of short blocks
that reacting long blocks must absorb before annihilating.
In the following, we shall show that each of the nested annihilations
in an actual simulation is correctly represented by the string replacements.

First, consider a subsequence starting with a long $0$-block and ending
with a long $1$-block with only short and just blocks between them.
Suppose further that the long blocks are long enough to
actually meet before turning into short or just blocks.
Then a maximal application of
Eqs.~(\ref{eq-firstcombine})--(\ref{eq-lastreact})
to the subsequence
subtracts the total ``shortness'' of the short
blocks from the long blocks yielding a diamond symbol,
which correctly represents what remains of the long blocks
when they meet.
The diamond will then annihilate according to
Eq.~(\ref{eq-diamondreact}) and yield the correct
number of extra groups and a symbol representing the
remainder (either a $\star$, if both long blocks are
completely annihilated, or a $0$ or $1$, if the
annihilation is partial).
This remainder symbol is exactly what would be the
result of an actual simulation of the subsequence
minus the just blocks that the sequence would
have emitted from both ends.
Because the just blocks do not affect long blocks outside
the subsequence, they can be disregarded.

Suppose then that either long block would decay
before meeting the other long block.
Then an application of Eqs.~(\ref{eq-firstdecay})--(\ref{eq-lastreact})
to the subsequence
will at some point turn either
the $0$- or $1$-symbol into a $\star$
(or the whole sequence can turn into a $0$, $1$, or $\star$).
A long block at one or both ends of the subsequence has thus turned
into a just or short block, making it possible for outside long
blocks to react over the remaining just and short blocks of
the subsequence.
Again, the just blocks that a decaying long block
would emit are disregarded as they do not
affect the length or order of other long blocks.

These are in fact all the cases that need to be considered,
as the above cases can be applied recursively to the
results of inner annihilations.
If neither case applies, we have found the final number
of groups, because either long $0$-blocks or long $1$-blocks
have then been exhausted and so there cannot be any other reactions
in an simulation nor can the replacements form any new $\Diamond$'s.
On the other hand, each of the applicable reactions
will be carried out at some point of a maximal application
of the replacements, because the $0$ and $1$-symbols
at the ends or a $\Diamond$ do not interact with outside symbols.
The replacements may evaluate the decay of the annihilating
long blocks in different order from the actual simulation,
but as the long blocks must absorb all short blocks
between them before they can meet and the short blocks
can not disappear unless absorbed by a long block,
the result is still the same.

Note that the remainder of a $\Diamond$ may be any
non-$\Diamond$ symbol, causing complicated recursion as the
result of an annihilation affects further replacements.
For example, consider the replacements
$$
\underbrace{\underbrace{0\star\underbrace{0\star1}_{\displaystyle1}
}_{\displaystyle0}\star\underbrace{0\star1}_{\displaystyle\star}1}_{
\displaystyle1},
$$
where the innermost reactions must be evaluated before
the outer reactions can be considered regardless of the
order of the possible replacements.

The iteration of
Eqs.~(\ref{eq-firstreact})--(\ref{eq-lastreact}) can be
carried out
by the stack algorithm shown in Fig.~\ref{fig-stackalg}.
Each symbol of the input string requires $O(l)$ operations
to be dealt with in this algorithm, where $l$ is the bit-length of the
symbol. This also includes the final steps to wrap the stack.
Therefore, the running time of this algorithm is
obviously $O(L)$ for a lattice of length $L$.

\ifreport
\figureIII{tbp}
\fi

The worst-case time for simply running the cellular automaton simulation,
on the other hand,
is $\Omega(L^2)$ since in this system far-away cars do interact with each
other.
For example, the initial state $0^{m+1}(1^k0^m)^{T-1}1^{k+1}$
(in the road representation)
would require $T$ simulation steps to reach a cyclic state.

Figure~\ref{fig-evolution-stack} illustrates an application
of the algorithm on an example string.

\ifreport
\figureIV{tbp}
\fi

\section{Steady-state conditions for the stack machine at the infinite limit}
\label{sec-exact-solution}

It is often the case that
infinite limits of systems are easier to solve than finite cases.
This is also the case with $\Rule m,k$: in this section, we
examine the behavior of the stack machine algorithm as the length
approaches infinity.
We regard the evolution of the stack configurations as a Markov process
and derive a set of equations
for the stationarity of the probabilities of stack configurations.
These equations are solved in the next Section
to determine the probabilities of different annihilations
and thus the steady-state flow.

Consider the symbols and reactions defined
above in Section~\ref{sec-symbol-def}.
All new groups are created from the diamonds,
which in turn can only arise when
a $0$-symbol is combined with a following $1$-symbol.
The algorithm in Fig.~\ref{fig-evolution-stack}
uses this fact to
find the final number of groups
by only tracking the reactions of $0$-symbols.
It scans through the input string linearly, from left to right,
maintaining
a stack of the processed symbols with all reactions
of $0$'s and $\Diamond$'s carried out.
This means that all the remaining $0$-symbols end up on top,
which we shall now call the {\em active part} of the stack.
When a non-$0$ input symbol consumes all the $0$'s,
we say that the symbol is dropped off
from the bottom of the active stack into the {\em inactive part},
as the symbol can no longer create new groups with the
\emph{following} symbols.

In finite systems with periodic boundary conditions,
the processing of the input string is divided into two parts.
First, the input string is scanned as above.
When all of the input has been read, the inactive part of
the stack, comprising of $1$'s and $\star$'s,
is reprocessed with the zeroes in the active part, since the
$1$'s on the bottom of the stack may react with the $0$'s on top,
producing new groups.
The relative effect of this wrapping diminishes
as the length approaches infinity:
it is easy show that
we can ignore all symbols dropped to the inactive part
in the infinite limit.

To obtain the limiting flow, we thus need to evaluate the
average number of new groups produced as the stack algorithm
processes a new symbol.
When new symbols are input, the active stack can either
grow infinitely or remain finite.
If the active stack remains finite,
the probabilities of different active stacks will
eventually reach a stationary distribution.
Once the stationary distribution is known,
it is straightforward to calculate the expected number
of new groups for a random input symbol.

If, on the other hand, there are not enough $1$'s and $\star$'s to
annihilate the $0$'s and the stack grows infinitely,
we can use the dualism property and consider
the thus finite stack of $1$'s instead of $0$'s.
It turns out later that we do not even have to consider this
dualism explicitly, as the symmetry of the equations
is restored in the next section.

Formally, we regard the evolution of the active stack
as a Markov process.
The
Markov property is satisfied as the next state depends only on the
current state and the upcoming independently distributed random symbol.
Clearly the process can reach each possible stack configuration
from every state and has a positive probability to remain in its current state
(the symbol $\star_{0,0}$).
Furthermore, given that the active stack will not grow infinitely,
the process will return to every state
an infinite number of times with probability one.
This means that the Markov chain is irreducible, aperiodic, and
Harris recurrent and therefore will converge to a unique stationary
distribution (see, e.g., \cite{FellerI} or
\cite[Proposition~6.3]{Nummelin1984}). %

An essential property is that the algorithm can be applied
independent of the lower levels for each stacked symbol
until that sub-stack is finished, that is, until the lowest level
of the sub-stack turns into a $1$ or a $\star$,
which happens either immediately,
if the stacked symbol is already a $1$ or a $\star$, or
when a $1$ or $\star$ on higher level falls to the bottom level
consuming all $0$'s on the sub-stack.
This allows us to consider each level of the active stack
as the bottom of an identically distributed sub-stack.
The distribution is particularly interesting
when a sub-stack is just finished.
At these times the active stack consists of zeroes
at the bottom of each sub-stack and of the $1$ or $\star$
that finishes the topmost sub-stack (see Fig.~\ref{fig-evolution-stack}).

Suppose $p(?_{x,y})$ is the distribution of symbols seen on the bottom
level of the active stack at each time a sub-stack is finished.
Then, if the symbol is a $0$, the same
distribution of symbols will be seen on the bottom of the sub-stack
above that symbol.
Thus, a symbol distribution defines a distribution
for stack configurations.
Note that an arbitrary stack
distribution can not be represented by such a symbol distribution
but it is required that the stack symbols are identically and independently
distributed and that the height of the stack is implied by the stack symbols
as described above.
Furthermore, even though each input symbol starts a sub-stack
so that there is the same number of time steps as there are input
symbols, the sub-stacks are finished out-of-synch with
the time steps of the algorithm.
This complication, however,
is inconsequential as the expected density of new groups on
each time step is still the same as the expected per-symbol density.

For simplicity, we consider the input symbol distribution $p_i(?_{x,y})$
to represent what remains of the symbol after initial
reactions of $\Diamond$-symbols.
This difference is only conceptual as the algorithm would
immediately carry out the initial reactions for each input symbol anyway.
It is easy to see that the resulting
distribution must still be geometric; only a constant is required to
normalize the lack of $\Diamond$'s. The normalized input distribution
is defined for indices in the set
$$
D := \{\,(x,y): x>-m,\ y>-k,\ [x>0][y>0] = 0\,\}
$$
as
\begin{eqnarray}
p_i(?_{x,y}) &:=& {A(1-\rho)^x \rho^y \over 1-A},
\end{eqnarray}
where we define the symbol $A$ to represent the quantity
\begin{eqnarray}
A &:=& (1-\rho)^m \rho^k,
\end{eqnarray}
which occurs often in the formulas below.
This quantity is
the initial probability of a $\Diamond$,
and the initial reactions
produce a density of $A/(1-A)$ new groups
(cf. Sec.~\ref{sec-simple-limits}).
Here and in the following, we use densities
relative to the initial symbol density $1$
in the string of symbols.

In our model the initial stack distribution is defined by
$p(\cdot|t=0) := p_i(\cdot)$ corresponding to the distribution of stacks
that results from running the algorithm on random input until the first
non-$0$ symbol is stacked in, i.e., at the first time a sub-stack
is finished (cf. Fig.~\ref{fig-evolution-stack}).

The transition from a symbol distribution $p(\cdot|t)$ to the
distribution $p(\cdot|t+1)$ on next time step can be defined
by considering the possible ways for a given symbol to arise
on each level of the stack: A symbol can be the result
of a $0$ and a non-$0$ above it reacting as per
Eqs.~(\ref{eq-zerostarreact}) and (\ref{eq-zeroonereact}).
The reaction will result in a non-$\Diamond$ symbol $?_{x,y}$
with probability
$$
p_c(?_{x,y}|t) =
\sum_{a<x;\,a\le0}\sum_{b\le0} p(0_{x-a,b}|t)p(?_{a,y-b}|t).
$$
If, on the other hand, the result of the reaction is a diamond,
it will further react, possibly several times,
according to Eq.~(\ref{eq-diamondreact}) and
yield $?_{x,y}$, where $(x,y)\in D$,
with probability
\begin{eqnarray*}
p_d(?_{x,y}|t) &=&
\sum_{r\ge1}\sum_{a\le0}\sum_{b\le0} p(0_{x-a+rm,b}|t)p(1_{a,y-b+rk}|t).
\end{eqnarray*}
In case there is another $0$ on top of a $0_{x,y}$, no reactions will occur
and the $0_{x,y}$ remains for the next time step. Finally, if the
symbol is not a $0$, it falls off from the bottom of the sub-stack and is
replaced by a fresh symbol from the input distribution.
Thus, the transition function can be written as
\begin{eqnarray}
&&p(?_{x,y}|t+1)\nonumber\\ &&\nonumber
\quad=\ [(x,y)\in D]p_d(?_{x,y}|t) + [(x,y)\in E]p_c(?_{x,y}|t) \\
&&\qquad +\ [x>0]p(?_{x,y}|t)p(0|t) + (1-p(0|t))p_i(?_{x,y}), %
\label{eq-transition}
\end{eqnarray}
where $p(0|t)$ denotes the total probability of $0$-symbols and
the set $E$ is defined as the complement of diamonds:
$$
E := \{\,(x,y): [x>0][y>0] = 0\,\}.
$$
Note that the
transition function does not define a Markov chain for symbols but it
implicitly defines a linear Markov operator for the subspace of stack
distributions that are determined by symbol distributions.

The stack distribution defined by a symbol distribution is stationary when
\begin{equation}\label{eq-balance}
p(?_{x,y}|t+1) = p(?_{x,y}|t).
\end{equation}
Thus, we need to find a symbol distribution $p(?_{x,y})$
corresponding to the unique stationary stack distribution.
The symbol distribution can then be used to determine probabilities
for different reactions.

\ifreport
\figureV{tbp}
\fi

Figure~\ref{fig-symbols} represents the possible indices for different
stack symbols. It is easy to see why some symbols can not occur at all, but
even more can be said. The distribution of the stack symbols retains
some of the geometric properties of the input distribution. The
distribution of $0$-symbols is geometric in its $x$ index and the
distribution of $1$-symbols is geometric in its $y$ index.
(The distribution of $?_{x,y}$ is in fact geometric in $y$
for all $(x,y)\in D$.)
This can be justified by noting that two
such symbols result from the same set of paths of the process
with the corresponding difference only in $x$ or $y$ index of
one specific input symbol (for $0$'s, the initial $0$ starting
the sub-stack and for $1$'s, the final $1$ that falls through to
the bottom of the sub-stack).
These properties are listed below: %
\begin{equation}\label{eq-symbol-geometric}
\begin{array}{rcll}
p(?_{x,y}) &=& 0, &x\le -m,\\ %
p(\Diamond_{x,y}) &=& 0, &x > 0, y > 0, \\
p(0_{x,y}) &=& (1-\rho)^{x-1} p(0_{1,y}), &x > 0, y \le 0, \\
p(1_{x,y}) &=& \rho^{y-1} p(1_{x,1}), &x\le0, y > 0.
\end{array}
\end{equation}
For a more rigorous proof, it is easy to check that the initial
distribution $p_i(\cdot)$ has these properties and that
the transition function maintains the properties.
Thus, the limiting stationary distribution must also have the
properties.

The stationarity recurrence given by Eqs.~(\ref{eq-transition}) and
(\ref{eq-balance}) can be transformed to
\begin{eqnarray}
{(1-[x>0]p(0))p(?_{x,y})\over(1-p(0))(1-\rho)^x \rho^y} = \nonumber
{[(x,y)\in D]p_d(?_{x,y})\over(1-p(0))(1-\rho)^x \rho^y}&&\\
+\ {[(x,y)\in E]p_c(?_{x,y})\over(1-p(0))(1-\rho)^x \rho^y}
+ {p_i(?_{x,y})\over(1-\rho)^x \rho^y}.&&\label{eq-balance2}
\end{eqnarray}
For clarity, we define $\g_{x,y}$ as the left side of the equation:
\begin{eqnarray}\label{eq-g}
\g_{x,y} &:=& (1-\rho)^{-x} \rho^{-y} {1-[x>0]p(0)\over 1-p(0)} p(?_{x,y}).
\end{eqnarray}
This transformation of $p(\cdot)$ cancels out the geometric factors and
will decouple the recursive $p(0)$ coefficient from the stationarity
equation.
The transformation is reversible as $p(0)$ can be obtained in terms of
$\g_{x,y}$ from
\begin{eqnarray}
p(0) &=& \sum_{x>0,y}p(0_{x,y})
= \sum_{x>0,y}(1-\rho)^x\rho^y \g_{x,y}.
\end{eqnarray}
With this definition, the geometric properties reduce to
\begin{equation}
\begin{array}{rcll}
\g_{x,y} &=& \g_{x,1}, & x\le 0, y > 0,\\
\g_{x,y} &=& \g_{1,y}, & x > 0, y \le 0.
\end{array}
\end{equation}
We define analogously the components $d_{x,y}$ and $c_{x,y}$
corresponding to terms on the right side of Eq.~(\ref{eq-balance2})
and apply the definition of $\g_{x,y}$ and the above properties:
$$
\begin{array}{l}
\displaystyle d_{x,y} := (1-\rho)^{-x} \rho^{-y} {p_d(?_{x,y})\over1-p(0)} \\
\displaystyle \qquad=
\sum_{r\ge1}A^r\sum_{a\le0}\sum_{b\le0} \g_{x-a+rm,b}\,\g_{a,y-b+rk}\\
\displaystyle \qquad={A\over1-A}\sum_{a\le0}\sum_{b\le0} \g_{1,b}\,\g_{a,1}
\end{array}
$$
for $(x,y)\in D$ and
\begin{eqnarray*}
c_{x,y} &:=& (1-\rho)^{-x} \rho^{-y} {p_c(?_{x,y})\over 1-p(0)} \\
&=& \sum_{a\le0}\sum_{b\le0} \g_{x-a,b}\,\g_{a,y-b}\\
&=& \sum_{a\le0}\sum_{b\le0} \g_{1,b}\,\g_{a,y-b}
\end{eqnarray*}
for $(x,y)\in E$.
The stationarity condition given in Eq.~(\ref{eq-balance2})
can then be expanded as
\begin{eqnarray*}
\g_{x,y} &=& [(x,y)\in D]d_{x,y} + [(x,y)\in E]c_{x,y}\\
&& \qquad\qquad\quad+\ (1-\rho)^{-x} \rho^{-y} p_i(?_{x,y})\\
&=& [(x,y)\in D]{A\over1-A}\sum_{a}\sum_{b} \g_{1,b}\,\g_{a,1}\\
&&+ [(x,y)\in E]\sum_{a<x;\,a\le0}\sum_{b} \g_{1,b}\,\g_{a,y-b} \\
&&+ [(x,y)\in D]{A\over 1-A},
\end{eqnarray*}
where we have left out the summation limits for zero terms,
based on $f_{a,b} = 0$ for $a>0$ and $b>0$.

Noting that only the middle line really depends on $x$ and
$y$ and that $\g_{a,\cdot} = 0$ for $a<-m$, the recurrence can be
written as
\begin{eqnarray}
\g_{x,y} &=& [(x,y)\in D]{A\over C} \nonumber\\
&&+ \sum_{a<x;\,a\le0}\sum_{b} \g_{1,b}\,\g_{a,y-b} \nonumber\\
&&-\ [x>0][y>0]\left({1-A\over C}-1\right),\label{eq-balance3}
\end{eqnarray}
where we define
\begin{eqnarray}\label{eq-C}
C &:=& {A \over \g_{1-m,1}}
= {1-A \over \displaystyle\sum_{a}\sum_{b} \g_{1,b}\,\g_{a,1} + 1}
\end{eqnarray}
for reasons to become clear later.

We have thus reduced the stationarity of the stack distribution
given by Eqs.~(\ref{eq-transition}) and (\ref{eq-balance}) to the
above convolution equation, where $f_{x,y}$ determines $p(?_{x,y})$ and
is given by Eq.~(\ref{eq-g}).

\section{Exact solution for the steady-state flow through generating functions}
\label{sec-genfunc}

The convolution recurrence in Eq.~(\ref{eq-balance3}),
which is the stationarity condition for the stack distributions,
can be solved using
generating functions (cf.~\cite{concmath}).
We define a formal generating function $\G_x(z)$ for the sequence
$\g_{x,\cdot}$ by
\begin{eqnarray}
\G_x(z) &:=& \sum_y \g_{x,y} z^y.
\end{eqnarray}
This generating function is not quite ordinary: the sum goes over all
$y$, positive and negative.
In general the values of a generating function do not
uniquely determine a sequence that is positive at an infinite distance
in both directions.
Here, however, we know that both $\G_1(z)$ and $\G_{1-m}(z)$ are
uniquely defined generating functions, because
when $x=1-m$,
the term $\g_{x,y}$ is only
positive at an infinite distance in the positive $y$ direction and
when $x=1$, it is only positive at an infinite distance
in the negative
$y$-direction (see Fig.~\ref{fig-symbols}).
The coefficients for the functions for other $x$ are
positive infinitely in both the positive and negative
$y$-direction but they are only
used in the following calculations
for formal multiplication and addition operations corresponding
to well-defined convolution and sum operations on sequences.

The term $\g_{x,y}$ given by Eq.~(\ref{eq-balance3})
vanishes for $x\le-m$. Thus, we can represent
it by the first non-zero case
$\g_{1-m,y} = [y\ge1-k]A/C$,
and the differences
\begin{eqnarray}
\g_{x,y} - \g_{x-1,y} &=& \nonumber
[x\le1]\sum_{b} \g_{1,b}\,\g_{x-1,y-b}\\
&&-\ [x=1][y>0]{1-C\over C}\qquad\qquad\label{eq-g-difference}
\end{eqnarray}
for $x>1-m$.
With the generating function notation we have
$\G_{1-m}(z) = (A/C) z^{1-k}/(1 - z)$
and
\begin{eqnarray*}
\G_x(z) &-& \G_{x-1}(z) = \G_1(z) \G_{x-1}(z)
- {1-C\over C}{[x=1]z\over 1-z}
\end{eqnarray*}
for $1-m<x\le1$. Thus,
\begin{eqnarray}\label{eq-gx}
\G_x(z) &=& \left(\G_1(z) + 1 \right)^{x+m-1} \G_{1-m}(z)
- {1-C\over C}{[x=1]z\over 1-z}
\end{eqnarray}
for $-m<x\le1$.
Thus, if we can solve $\G_1(z)$ and $C$,
we have determined $\G_x(z)$ for all $x$,
because $C$ determines $\G_{1-m}(z)$.

For $x=1$, equation~(\ref{eq-gx}) can be written as
\begin{eqnarray}\label{eq-g1}
\left(\G_1(z) + 1\right) &=&
\left(\G_1(z) + 1\right)^m
{A z^{1-k}\over C(1-z)} - {z - C\over C(1-z)},
\end{eqnarray}
which is an $m$th degree equation with respect to $\beta
:= \G_1(z) + 1$. It is easy to see that there are at most two
positive real solutions for $\beta$.

Because we can solve $\G_1(z)$ given $C$,
the complete solution for the stationary distribution of the stack
configurations and thereby $\phi$
now hinges on determining $C$.
Unfortunately,
the quantity $C$ cannot be solved directly from the above equations,
since its definition is already used in solving them;
equations relating $C$ and $\G_1(z)$ reduce to identities
when combined with Eq.~(\ref{eq-g1}).

However, there is a different, strange approach:
we can determine $C$
based on the fact that $\G_1(z)$ represents (indirectly)
a probability distribution.
The correct $\G_1(z)$ must obviously
be analytic for some region $|z|>r$.
Additionally,
it must be positive and decreasing in $z$, because it has non-zero
coefficients only for non-positive exponents of $z$, and
all coefficients must be nonnegative since they are probabilities
multiplied by a positive function of the index.
These two constraints allow us to uniquely determine $C$ in the following.

Eq.~(\ref{eq-g1}) can be solved with respect to $C$ as
\begin{eqnarray}
C &=& {\beta^m A z^{1-k} - z\over (1-z)\beta - 1}
\end{eqnarray}
(note that $\beta = \G_1(z) + 1$ depends on $z$).
Substituting $\alpha := 1/z$ in this equation yields a perfectly symmetric form
\begin{eqnarray}\label{eq-g-symmetric}
C &=& {A \alpha^k \beta^m - 1\over \alpha\beta - \alpha - \beta}.
\end{eqnarray}
Figure~\ref{fig-g-loop} depicts the solutions of Eq.~(\ref{eq-g-symmetric})
with different values of $C$ for $\Rule 2,2$ and $\Rule 3,2$.
The figures are essentially similar for larger $m$ and $k$ with
at most two positive solutions and in addition one everywhere negative
solution if $m$ is odd.
In either case, it can be seen that a too large value of
$C$ results to a gap in the solution and a too small value results to
either a non-monotonous or a non-positive solution. Only the correct
$C$ allows changing branches in the singularity point to obtain a
feasible solution. This is analogous to the singular behavior of
elliptic curves (cf.~\cite{elliptic}).

\ifreport
\figureVI{tbp}
\fi

Because the surface $C(\alpha, \beta)$ is smooth, the two constant-$C$
contours can only cross at a critical point.
The critical
points $\nabla C(\alpha,\beta) = 0$ are determined by the equations
$$
\left\{\begin{array}{ll}
A k \alpha^{k-1} \beta^m (\alpha\beta-\alpha-\beta)
- (A \alpha^k \beta^m - 1)(\beta - 1) = 0,\\
A m \alpha^k \beta^{m-1} (\alpha\beta-\alpha-\beta)
- (A \alpha^k \beta^m - 1)(\alpha - 1) = 0.
\end{array}\right.
$$
Multiplying by $(\alpha-1)$ and $-(\beta-1)$ and adding, the equations yield
$k(\alpha-1)\beta = m(\beta-1)\alpha$.
Changing variables to $a = (\alpha-1)/(m\alpha)$ and $b =
(\beta-1)/(k\beta)$ yields a simple form $a = b$ for the equation.
Substituting $\alpha = 1/(1-am)$ and $\beta = 1/(1-ak)$ to the critical point
equations and to Eq.~(\ref{eq-g-symmetric})
results in
\begin{equation}\label{eq-critical2}
\left\{\begin{array}{l}
\displaystyle
A \left(1-am\right)^{-k} \left(1-ak\right)^{-m}
= {a\over 1-a(k+m-1)}, \\
\displaystyle
A \left(1-am\right)^{-k} \left(1-ak\right)^{-m}
= 1 - C{1-a(k+m)\over(1-ak)(1-am)}.
\end{array}\right.
\end{equation}
Now, we only need to reduce $a$ from this system and
then we have an equation relating the unknown $C$ to the parameters
$A$, $m$, and $k$.
By eliminating the left sides we obtain from the right sides
either $a=1/(k+m)$ corresponding to a pole of Eq.~(\ref{eq-g-symmetric}) or
\begin{equation}\label{eq-quadratic-a}
C[1-a(k+m-1)] = (1-ak)(1-am).
\end{equation}
This is a second degree equation for $a$ and its smaller solution is
\begin{equation}\label{eq-a}\label{eq-first-solu}
a = {1 + (1 - C)(k + m - 1) - c \over 2km},
\end{equation}
where we define
\begin{equation}
c := \sqrt{\left(1 + (1 - C) (k + m - 1)\right)^2 - 4(1 - C)km }.
\end{equation}
Note that the other solution with $+c$ in Eq.~(\ref{eq-a})
does not yield a critical point.

Now Eq.~(\ref{eq-quadratic-a}) can be used to rewrite
the first equation of Eq.~(\ref{eq-critical2}) as
\begin{eqnarray}
A &=& C a \left(1-am\right)^{k-1} \left(1-ak\right)^{m-1}\\
\label{eq-last-solu}%
&=& C^k a \left(1-a(k+m-1)\right)^{k-1} \left(1-ak\right)^{m-k},
\end{eqnarray}
yielding $A$ as a function of $C$ in closed form
with the solution of $a$ given by Eq.~(\ref{eq-a}).
In the special case of $m=k$, we can solve $\rho$ from $A$
in closed form and obtain the density as a function of flow.

For any given $m$ and $k$, the
equations~(\ref{eq-first-solu})--(\ref{eq-last-solu}) for $A$ and $C$
can be expanded to a polynomial equation, which is easily seen to be
second degree in $A$ and at most $2(m+k)$th degree in $C$.
In practice the degree of $C$ seems to reduce to $m+k+1$,
and we conjecture that this holds for all $m$ and $k$.
For example, in the case of $m=k=2$, the equation can be written
out as
$$
16A^2 + 8AC^2 - 36AC^3 + (1 + 27A)C^4 - C^5 = 0.
$$

Now that we have determined all variables, we can determine
probabilities for different reactions. The density of new groups on
the bottom level of the active stack is
\begin{eqnarray*}
\rho_{\Delta g}
&=& \sum_{(x,y)\in D}\sum_{r\ge1}r\sum_{a\le0}\sum_{b\le0}
p(0_{x-a+rm,b})p(1_{a,y-b+rk})\\
&=& (1-p(0))\!\!\!\!\sum_{(x,y)\in D}\sum_{r\ge1}
rA^r\!\!\sum_{a,b\le0}\!\!(1-\rho)^x \rho^y \g_{1,b}\,\g_{a,1}\\
&=& (1-p(0)){A\over(1-A)^2}{1-A\over A}\left({1-A\over C}-1\right).
\end{eqnarray*}
When all levels of the stack are taken into account, the
total density of new groups is
$$
\rho_{\Delta g}(1 + p(0) + p(0)^2 + \ldots)
= {1\over C}-{1\over 1-A}.
$$
Adding in the initial density $1$ and the density of groups arising
from the initial reactions yields the final group density relative
to the initial group density:
$$
{\rho_G\over\rho(1-\rho)}
= 1 + {A \over 1-A} + \left({1\over C}-{1\over 1-A}\right)
= {1 \over C}.
$$
Thus, in the intermediate phase, $\phi = C$.
Furthermore, the
phase transitions occur where $C$ as a function of $\rho$ crosses
$m\rho$ or $k(1-\rho)$, the flow of free-flowing and congested phases.
For example, when $m=k=2$ the phase transitions can be solved to be exactly at
$$
\rho={1\over2}\pm{1\over7}\left(2\sqrt{2}-{5\over 2}\right).
$$

The most important results are summarized below:
\begin{prop}
The flow at infinite time for random infinite strings
in the intermediate phase
is $C$, where $1/C$ is the expected number of extra groups arising
per each input symbol and is determined by
Eqs.~(\ref{eq-first-solu})--(\ref{eq-last-solu}).
The flow depends on $\rho$ only through the
quantity $A = (1-\rho)^m\rho^k$.
\end{prop}
\begin{prop}
For any given $m$ and $k$, the flow $\phi$ at infinite time
and the density $\rho$ in the intermediate phase
can be related by a polynomial equation maximally
of degree $2(m+k)$.
\end{prop}

\section{Upper limit for steady-state flow in finite systems}
\label{sec-finite}

Carrying out calculations for finite systems is considerably more difficult,
since the probabilities are no longer independent of each other.
However, the following interesting limit can be derived.

\begin{prop}\label{prop-finite-max}
The average flow in steady states starting from random binary strings
of length $L$ with $0 < \rho < 1$ satisfies
$$
\overline \phi \le
\min\left(m \rho, 1 - {L \choose \rho L}^{-1},
k (1-\rho)\right),
$$
where equality applies at least when $k \ge L-1$ and $m \ge L-1$.
\end{prop}

Proof.
If $0 < \rho < 1$, the number of different initial states
with a given number of groups can be counted by considering
different ways of placing the group boundaries on the string.
The distribution of $N_G$ for random binary strings can be
simplified to
$$
p(N_G) =
\left\{
\begin{array}{r} \displaystyle
{\displaystyle L \over \displaystyle N_G} { \displaystyle
{\rho L-1 \choose N_G-1} {(1-\rho)L-1
\choose N_G-1}
\over
\displaystyle
{L \choose \rho L} } \mbox{, if $N_G>0$,} \\
0 \mbox{, if $N_G \le 0$.}
\end{array}
\right.
$$
Using basic binomial coefficient sum formulas (see e.g.~\cite{concmath})
and noting that $N_G$ cannot be zero,
the expected value of $L/N_G$ is
$$
\left< { L \over N_G } \right> =
\sum_{N_G} N_G p(N_G) =
{1 \over \rho (1-\rho)}
\left( 1 - {L \choose \rho L}^{-1} \right).
$$
From this, the formula in the Proposition follows. Finally, the equality
follows simply from the fact that if the condition given holds, no groups
can split.

Note that with reasonable $L$ and $\rho L$, the reciprocal of the binomial
coefficient is negligibly small compared to 1.

\section{Simulations}

Simulations were carried out to test the theoretical results.
For small $L$, complete summations were possible so the simulated
curves are in fact exact. For large $L$, a number of random initial
configurations were generated and the evolution of the system
simulated. Since the resulting
steady-state flow under $\Rule m,k$, when $m>1$ and $k>1$,
depends on the whole initial state (and not just $\rho$,
as for when $m=1$ or $k=1$), the samples so simulated will in general
not have the same flow. Therefore, the standard deviation is displayed
along with the average of the resulting flows, giving an idea of
the strength of the fluctuations. As $L$ tends to infinity, the
fluctuations slowly average out,
displaying the usual $1/\sqrt L$ behavior for the standard deviation.

Figure~\ref{fig-infinite} depicts the simulated flow and
exact solution for infinite space. The theoretical solution
agrees well with simulated results.
Figure~\ref{fig-finite} shows the simulated flow and
upper and lower limits for finite space.
For $m,k\ge L-1$ the upper limit is exact as confirmed
by the simulation.

\ifreport
\figureVII{tbp}
\figureVIII{tbp}
\fi

\section{Conclusions}

In this article, we have solved the behavior of the generalized
traffic rules $\Rule m,k$ for infinite lengths of road and uniform
random initial conditions.

We have derived an efficient algorithm for computing
the average flow from an initial state under the
generalized traffic rules $\Rule m,k$.
The idea behind the algorithm is an appropriate
representation of the road as a string of $0^x1^y$ blocks
instead of single sites, and the fact that finding the
average flow can be reduced to finding the number of
these blocks in the cyclic state.

The algorithm works by decoupling the time from the
simulation and considering directly the different reactions
that would happen during the evolution of the system.

Analysis of the algorithm in the infinite limit
yields an exact solution for the flow in an infinite space.
Simulated results agree perfectly with the analytic
solution.

Finite-space behavior is more complex because
single sites are no longer independent.
We have, however, been able
to obtain for the average flow a non-trivial upper limit,
which is exact for $m,k\ge L-1$.

\section*{Acknowledgments}

The authors would like to thank Rauli Ruohonen for discussions and comments
on this manuscript.

\ifreport
\def\pre{Phys.~Rev.~E}
\bibliographystyle{plain}
\fi

\end{document}